\documentclass{aa}
\usepackage{psfig}
\def\teff{$T\rm_{eff}$ }
\def\lambo{$\lambda$ Boo }
\def\$v \sin i ${v\,sin\,i}
\def\kms {$\mathrm{km\, s^{-1}}$}
\usepackage{graphics}
\usepackage{psfig}
\begin{document}
\title{
Spectra of binaries classified as $\lambda$ Boo stars
\thanks{Based on data from ESO, Observatoire du Pic du Midi, IUE
Final Archive and on observations obtained with the Hipparcos satellite.}  }
\author{R. Faraggiana\inst{1},
M. Gerbaldi\inst{2,3},
P. Bonifacio\inst{4} and 
P. Fran\c cois\inst{5,6}}                                

\offprints{R. Faraggiana}
\institute{Dipartimento di Astronomia dell' Universit\`a di Trieste,
Via G.B.Tiepolo 11, I-34131 Trieste, Italy \\
email: faraggiana@ts.astro.it
\and Institut d'Astrophysique, 98 bis Bd Arago, F-75014 Paris, France \\ 
email: gerbaldi@iap.fr
\and
Universit\'e de Paris Sud XI
\and
Istituto Nazionale per l'Astrofisica - 
Osservatorio Astronomico di Trieste, 
Via G.B.Tiepolo 11, I-34131 Trieste, Italy \\
email: bonifaci@ts.astro.it
\and
European Southern Observatory(ESO), Alonso de Cordoba 3107, Vitacura, 
Casilla 19001, Santiago19, Chile \\
email:fpatrick@eso.org
\and
Observatoire de Paris-Meudon, DASGAL, 5 Place Jules Janssen, F-92195 Meudon, 
France \\
e-mail: Patrick.Francois@obspm.fr                     
}
\mail{faraggiana@ts.astro.it}
\date{received .../Accepted...}
\authorrunning{R. Faraggiana et al}

\abstract{
High angular resolution observations have shown that some stars 
classified as \lambo are binaries 
with low values of angular 
separation and magnitude difference of the components;  therefore the observed 
spectrum of these objects is 
a combination of those of the two components. 
These composite spectra have been used to define
spectroscopic criteria able to 
detect other binaries among stars classified as \lambo.
The application of this method to HD 111786 is presented: the contribution
of 5 components to the observed spectrum is demonstrated by the shape 
of the O I 7774 \AA~ feature.
This result makes unreliable any attempt to perform an abundance analysis
of this object which therefore must be definitely rejected from the class 
of the peculiar \lambo stars. 
This approach allowed us also to recognize that the SB2 star HD 153808 is 
in reality  a triple system. 
} 
\maketitle{}
\section{Introduction}

Metal underabundances are not common among young A-type stars; following
the description of the \lambo spectrum by Morgan et al (1943),
Burbidge \& Burbidge (1956) made a first abundance analysis of
other metal-weak stars which are now known to belong to both \lambo 
and field HB classes. 
Sargent (1965a and 1965b) was the first to
list the criteria to select the Population I early A-type \lambo stars.
Later on, Slettebak et al (1968) gave a more detailed spectroscopic 
definition as well as the criteria to distinguish \lambo stars from other 
metal weak stars. 
One of the specified characteristics is the moderately large rotational
velocity (usually 100-150 \kms ) which clearly represented a difficulty
for abundance analyses performed with the curve of growth method and
based on photographically recorded spectra. In fact, the only detailed study 
of the abundance peculiarities before the advent of modern detectors is 
that by Baschek \& Searle (1969).

Systematic spectroscopic and photometric search for new members of the 
class have been performed mainly in the past two decades.
The spectroscopic characteristics  in classification
dispersion covering the optical range are extensively described by Gray (1997)
who gave a new definition and more detailed classification criteria to 
select these objects.
  
The comparison of the Gray's definition with the earlier ones shows that
the new definition includes also later 
spectral types up to early F-type and stars with low $v \sin i $ values.

Modern detailed abundance analyses are still scanty and concern a very limited 
number of objects (e.g. Venn \& Lambert, 1990; St\"urenburg, 1993), 
but sufficient to note that the metal 
deficiency is widely variable from
star to star with a large scatter of pattern behaviour.
All the abundance analyses (with the exception of that on 
HD 84948, see below) are based on the hypothesis that \lambo
stars are single objects. In this paper we investigate
spectroscopic criteria to detect duplicity among these stars.

In the meantime theories to explain the \lambo phenomenon have been 
proposed, but none is definitely accepted. In fact, the \lambo
phenomenon is complicated; for example 
several stars show one or more
of the following  peculiarities:
an IR excess, photometric variability,
short term pulsation or narrow absorption lines,
indicating the presence of a shell.  Some stars instead
show none of the above.

Faraggiana \& Bonifacio (1997) proposed a new interpretation of these 
objects; 
they analyzed 
the combined list of \lambo candidates extracted from  three different 
and recent sources 
and noticed that a considerable number of these \lambo candidates are 
detected or suspected binaries.
In the present paper we discuss how it is possible to recognize a
composite spectrum of a moderately high rotating star ($v \sin i $ up to
about 100 \kms) 
through the comparison of observed and computed spectra.
In fact high resolution combined with low noise spectra are quite easily
obtainable with modern equipment and allow one to detect a number of previously
unsuspected binaries.

A search for spectral features which can be used as duplicity criteria for
A-F stars can find application on a much wider field than that of stars more
or less questionably classified as \lambo.

The stars used to select spectroscopic  
signs of duplicity are chosen among those
found to be SB2 spectroscopic binaries or binaries with a separation less
than 1 arcsec measured by speckle interferometry 
and/or extracted from the Hipparcos Catalogue (ESA, 1997) and a magnitude 
difference of less than 1.5 mag.
The spectra of 5 binaries, 4 of which classified as \lambo stars (HD 38545,
HD 47152, HD 84948 and HD 153808) by at least one author and one 
(HD 18622) classified as normal star, 
are discussed and the spectroscopic criteria that reveal their duplicity 
investigated.

A first application to HD 111786, classified as \lambo by Gray 
(1988), is given in this paper; 
Faraggiana et al (1997) have discovered 
a companion through the presence of
the narrow lines previously interpreted as originating from a shell 
around a single  rapidly rotating object.
The present analysis of the O I triplet at 7774 \AA~ in particular
demonstrates that in reality HD 111786 is formed by a clump of stars with 
similar luminosity and therefore any classification and abundance analysis
of this object must be considered unreliable.

\begin{figure}
\psfig{figure=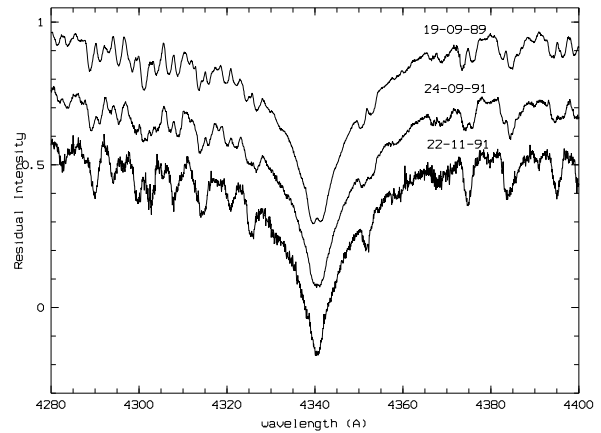,width=8.8cm,angle=-90}
\caption{The spectrum of the SB2 star HD 18622 in the region of H$_{\gamma}$ 
at different dates.}
\label{fig1}
\end{figure}

\begin{figure}
\psfig{figure=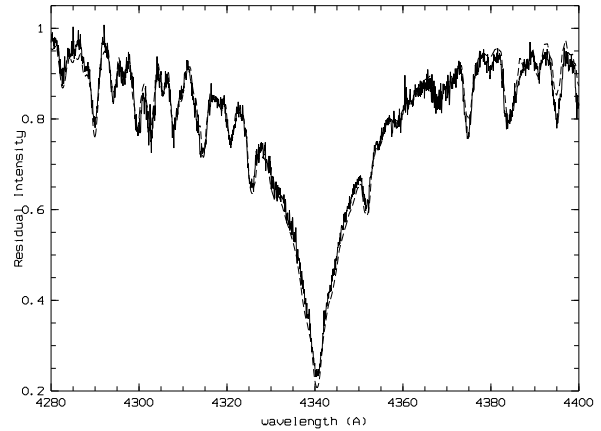,width=8.8cm,angle=-90}
\caption{The spectrum of HD 18622 taken on 22 Nov 1991 superimposed on 
the computed one (dashed line).}
\label{fig2}
\end{figure}

\begin{figure*}
\psfig{figure=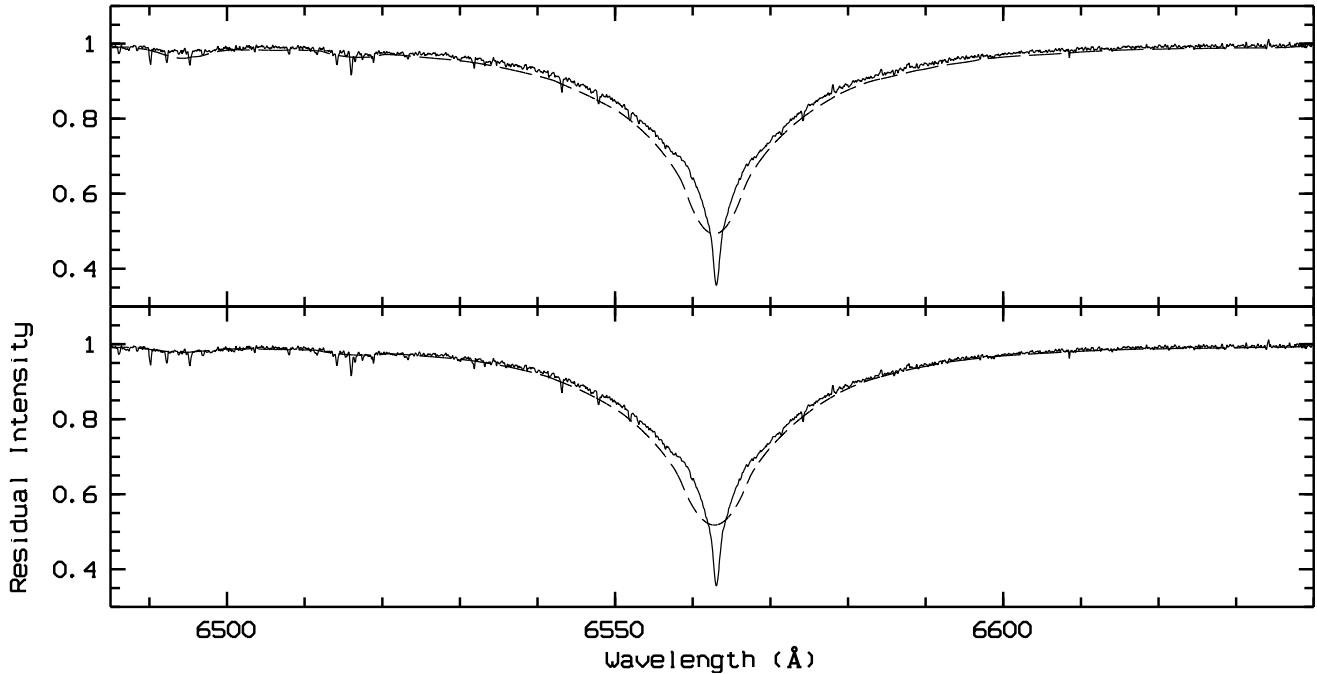,clip=t}
\caption{The observed spectrum of HD 38545 in the region 
of H$_{\alpha}$ compared with two synthetic spectra (dashed line): 
that computed with the parameters
given in Table 3  (top: \teff=8500 K, log~g=3.6, $v \sin i$
=175 \kms ) 
and that computed with the St93 parameters  (bottom: \teff=9000 K, 
log~g=3.6, $v \sin i$=200 \kms ). }
\label{fig3}
\end{figure*}

\begin{figure}
\psfig{figure=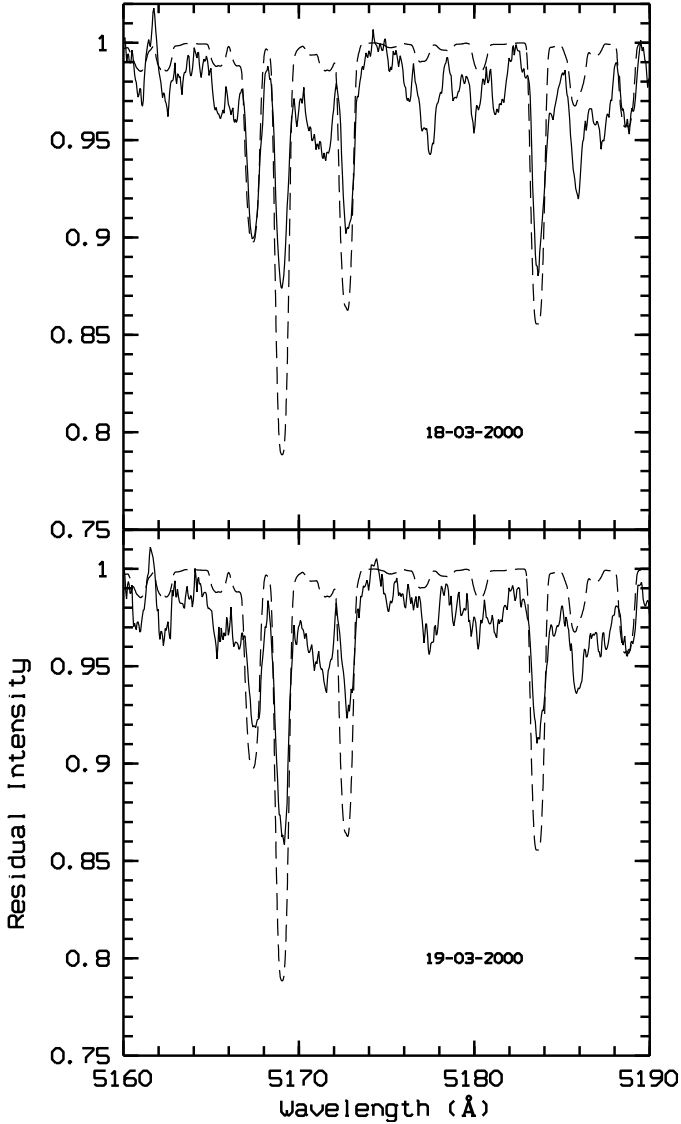,clip=t}
\caption{The spectrum of HD 47152 in the region 5160-5190 \AA~ shows the  
poor fit with the computed spectrum (dashed line), the high number of 
unidentified lines and
the variability of the Mg I triplet in 1 day.}
\label{fig5}
\end{figure}

\begin{figure}
\psfig{figure=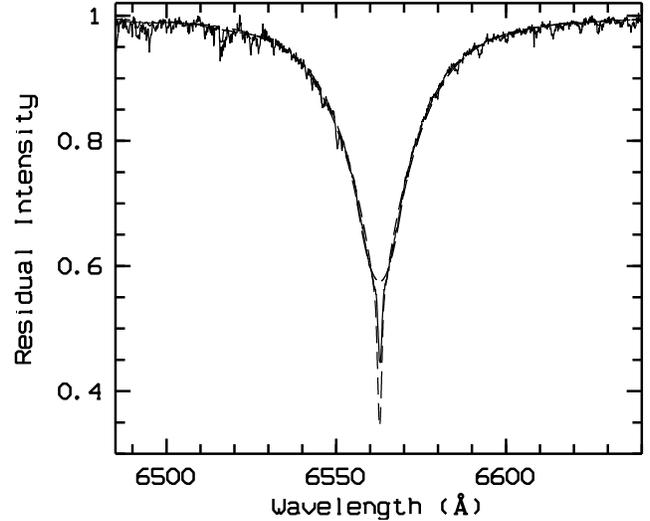,clip=t}
\caption{ The H$\alpha$ profile of HD 47152 compared with two synthetic 
spectra  both computed with \teff = 10000 K, log g= 4.0 and two
values of the broadening  30 and 300  \kms .
}
\label{fig6}
\end{figure}

\begin{figure}
\psfig{figure=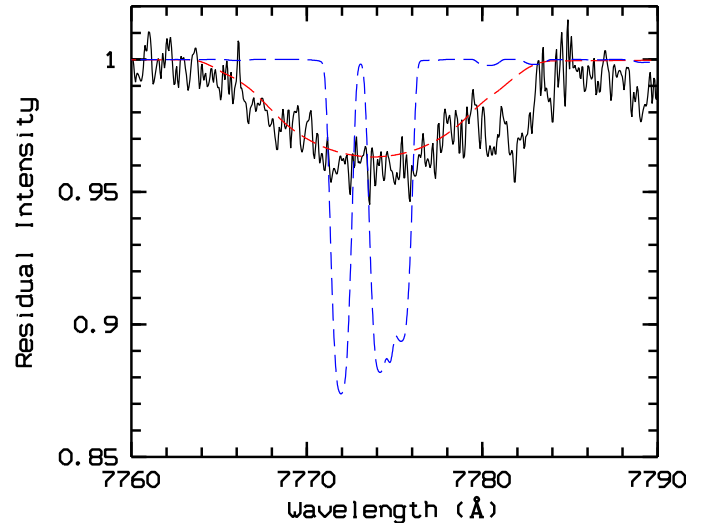,clip=t}
\caption{ The O I triplet 7772-7775 is not resolved in the spectrum of HD 
47152, as it should
be expected from the other narrow metal lines; The computed spectra (dashed
lines) have been broadened by 30 \kms and 300 \kms respectively.}
\label{fig7}
\end{figure}

\begin{figure}
\psfig{figure=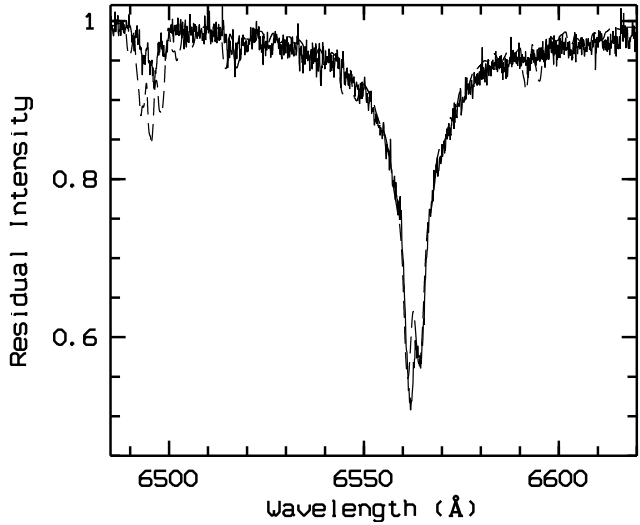,clip=t}
\caption{The observed H$_{\alpha}$ profile of the SB2 star HD 84948
compared with the spectrum (dashed line) 
obtained by combining two computed spectra 
(see text).}
\label{fig8}
\end{figure}

\begin{figure}
\psfig{figure=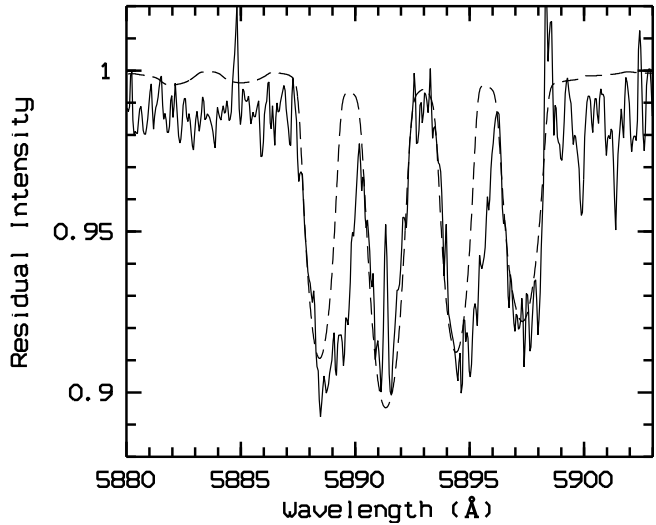,clip=t}
\caption{The region of the Na I doublet of HD 84948 compared with the
spectrum computed with the same combination of models used for Fig.
 \ref{fig8}.} 
\label{fig9}
\end{figure}                  
            
\begin{figure}
\psfig{figure=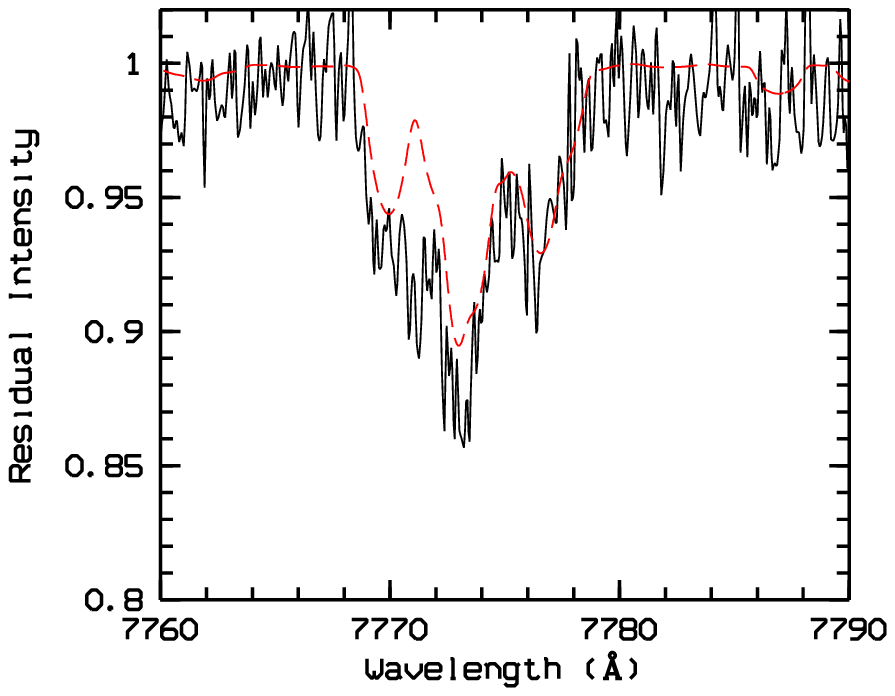,clip=t}
\caption{The region of the O I triplet of HD 84948 compared
with the same combination of spectra used for figures 7 and 8
(see text).} 
\label{fig8bis}
\end{figure}

\section{
Classification and 
duplicity data for the selected sample of binaries}

The five already known binary stars which have been selected for the search of 
duplicity signatures in their spectrum are listed in Table 1. The V 
magnitudes and the spectral types are taken from the Bright Star Catalogue
(Hoffleit \& Warren, 1994); the parallaxes and errors, measured in milliarcsec 
(mas) and the Hipparcos magnitudes (Hp) are from the Hipparcos  
Catalogue (ESA, 1997). The angular separations are from the same source
for all but the last star; the value for HD 153808 is that given in the 
Washington Visual Double Star (WDS) catalogue (Worley \& Douglass, 1997).

The sources of  information
for the spectroscopic binaries are
Corbally (1984) for HD 18622, Paunzen et al (1998) for HD 84948, 
and Petrie (1939) for HD 153808.

\bigskip

\begin{table*}
\caption{Observational data; V and spectral type are from the Bright Star 
Catalogue (Hoffleit \& Warren, 1994)
for all but HD 84948 whose values are taken from the CDS database;  in 
Column 9  the authors who classified as \lambo are given (G= Gray, 1997; CC=
Paunzen et al, 1997; AM= Abt \& Morrell, 1995) for all but the first star.
                               }
\begin{center}
\begin{tabular}{cccrcclcccc}
\hline

HD    & V    & Sp.    Type  & $\pi$   & $\sigma(\pi)$ & Hp(A) & Hp(B) & Sep. & Class. & Rem.   \\

      &      &   BSC        &  mas    &  mas        &         &       & arcsec & & & \\            
\hline

18622 & 3.24 & A5 IV & 20.22 & 0.54 & 3.278   & 4.423 & 8.310 &  st. $v\sin i$ & SB2\\

38545 & 5.72 & A3 Vn  & 7.72  & 0.93 & 6.229   & 6.87  & 0.155 & Gray, CC & speckle \\

47152 & 5.79 & B9npEu  & 7.74  & 1.08 & 6.201   & 6.969 & 0.210 &  AM & speckle\\

84948 & 8.13 & F0    & 4.97  & 1.14 & --      & --    & --    & CC & SB2 \\

153808& 3.92 & A0 V   & 20.04 & 0.65 &         &       & 0.2   & AM & SBO \\
\hline
\end{tabular}
\end{center}
  \label{ Table 1
         }

\end{table*}

{\bf HD 18622} Non peculiar star, classified A3 V (Bright Star Catalogue, 
1964 ed), A4 III (CDS) or A5 IV (Gray \&
Garrison 1989).
This  star belongs to the catalog of $v \sin i $ standard stars by
Slettebak et al (1975) and to the list of standard stars for 
H$_{\alpha}$ photometry by Strauss and Ducati (1981). 
The star has a visual companion, HD 18623, at 8.3 arcsec with a 
${\Delta}$m=1.1; so
it is an example of a spectroscopic 
binary with a more distant visual companion.
The only references to its SB2 nature 
we aware of
are in the Bright
Star Catalogue (since the 1982 ed.),  and in Corbally (1984).

{\bf HD 38545} Classified \lambo by Gray \& Garrison (1987); 
the presence
of shell lines has been noted first by St\"urenburg (1993) and since then 
studied both in the visual (Bohlender \& Walker, 1994; 
Andrillat et al, 1995;
Hauck et al, 1995 and 1998; Holweger \& Rentzsch-Holm, 1995; Holweger et 
al, 1999; Hauck and Jaschek, 2000) and in the UV range (Grady et al, 1996). 
The binary nature 
of this object 
has been discovered by McAlister et al (1993)
and confirmed by Hipparcos (ESA, 1997) and new speckle  
data (Marchetti et al, 2001). 

According to the Bright Star Catalogue (since its 1982 ed.), 
this star has a variable radial velocity whose 
origin has never been studied.
 
{\bf HD 47152} Classified either Ap of  Hg,4077 type (Osawa, 1965)
or \lambo (Abt \& Morrell, 1995), but 
excluded from the \lambo class by Hauck et al (1998) who recall
the agreement of the Osawa classification with the Geneva ${\Delta}$(V1-G)
parameter. 
In spite of the numerous speckle observations 
since 1982 indicating the presence of a companion 
(McAlister \& Hendry, 1982; Bonneau et al 1984; etc.), the duplicity 
is ignored in all the spectroscopic studies of its peculiarities.

The values of the angular separation and the magnitude difference given 
by Hipparcos,  
are such that only a composite spectrum can be observed; so the star
cannot be safely assigned either to the Ap class or to that of \lambo stars
by neglecting its duplicity.

{\bf HD 84948} The classification  as possible \lambo is due to Abt 
(1984);
Andrillat et al (1995) on the basis of near-IR spectra classified it as \lambo 
with shell; Paunzen and Gray (1997) note the similarity with 
HD 84123 
spectrum and suggest that both are perhaps field HB stars; Paunzen et al 
(1998) performed an abundance analysis based on an 
echelle spectrum of this resolved SB2 
object (range 4000-5700 \AA ), extended to Na through unquoted 
observational data.

{\bf HD 153808} Intriguing object for which there are discordant results on its 
duplicity as a visual binary. 
Hipparcos did not detect any sign of duplicity 
nor is its  duplicity  mentioned in the Hipparcos Input
Catalogue (Turon et al, 1993).
Controversial visual binary detections are reported in the literature,
the two values of the separation 0.24 and 0.97 arcsec measured by Isobe 
et al (1990 and 1992), being not 
confirmed by other authors.

As a spectroscopic binary this object has been studied in the past; the 
period, P=4.0235 d as well as the other
orbital elements are given in the Batten et al (1989) catalogue, but
are based only on the measures by Luyten (1936).
The double-lined  spectrum is discussed by
Petrie (1939) who classified the two components A0 and A2 and
computed a magnitude difference of 1.5. 

It is classified also as Mn star (e.g. Wolff \& Preston, 1978), or normal
A0IV$^{+}$ by Gray \& Garrison (1987) and quoted as SB2 by Wolff (1978). 

The values of the projected rotational velocity vary considerably from author
to author: 38 \kms~ (Lambert et al, 
1986); 50 \kms~ (Abt \& Morrell, 1995); 80 \kms ~
(Wolff \& Preston 1978); 90 \kms ~ (Slettebak, 1954).

It is included in the Tokovinin (1997) catalogue of triple systems; this 
author considers doubtful the
binary detection of the visual AB system since it is not confirmed 
by the Hipparcos data, but retains the spectroscopic duplicity Aab.

\section{Observations}

The optical spectra have been collected at the Echelec (ESO), Musicos 
(Observatoire du Pic du Midi) and FEROS (ESO)
spectrographs which have a nominal resolution of about 28000, 38000 and 48000 
respectively; the wavelength range of each of them is broad
enough to include at least one Balmer line. 
The journal of observations is given in Table 2.

\begin{figure*}
\psfig{figure=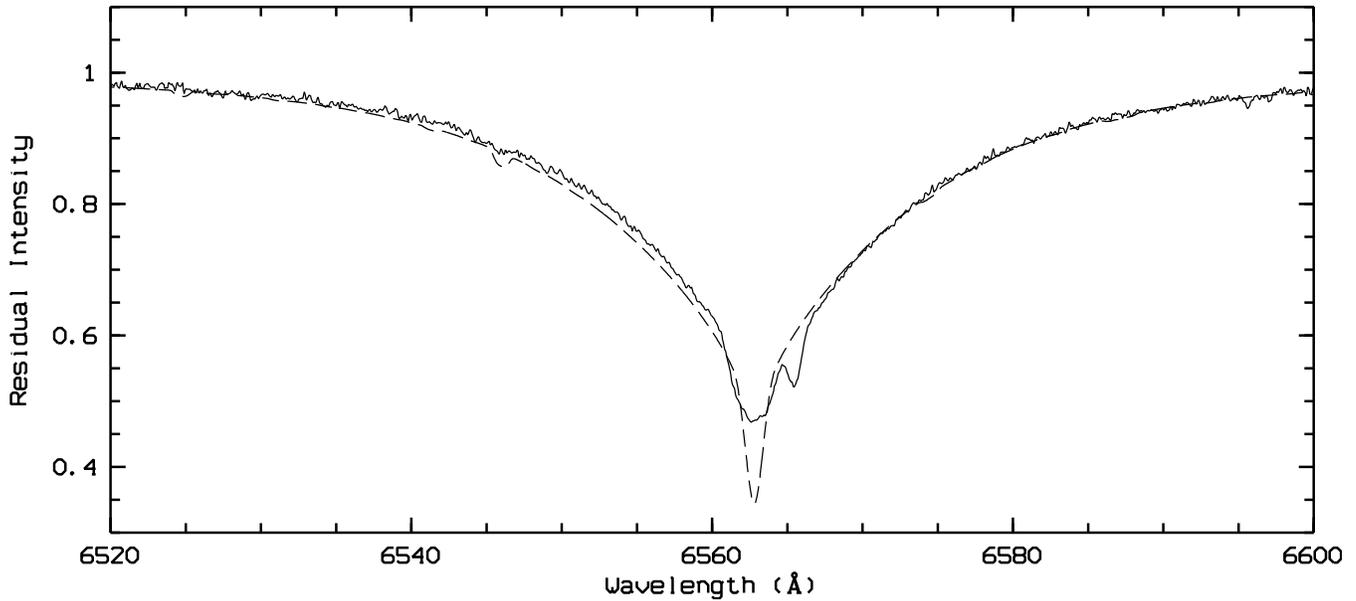,clip=t}
\caption{
The observed profile of H$_{\alpha}$ of HD 153808 compared with that 
computed from a model with \teff=10000 K, log g= 4, broadening=30 
\kms .}
\label{fig10}
\end{figure*}

\begin{figure}
\psfig{figure=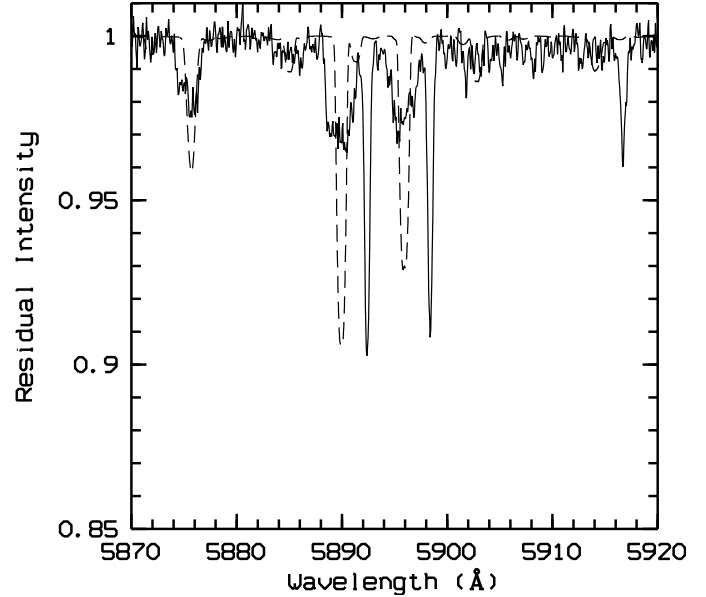,clip=t}
\caption{
The duplicity of the blue-shifted component star of HD 153808 is clear from 
the Na I doublet and the He I 5876  profiles compared with the computed 
spectrum (broken line) from the same model used for Fig. \ref{fig10}. }
\label{fig11}
\end{figure}

\begin{figure}
\psfig{figure=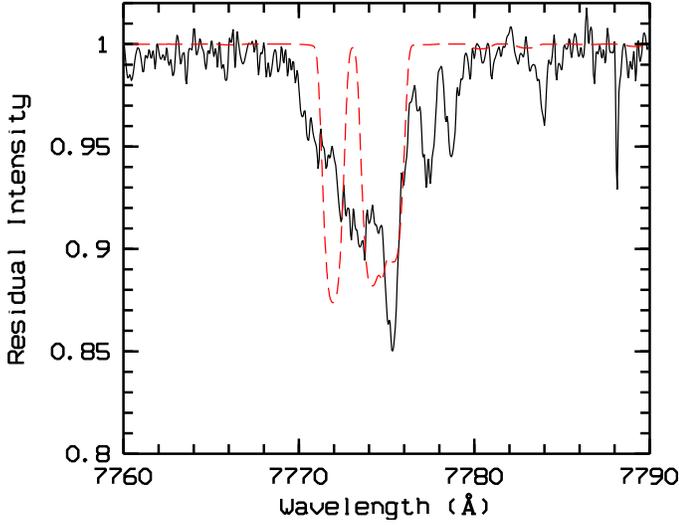,clip=t}
\caption{The region of the O I triplet of HD 153808 compared
with the spectrum computed using the model used in figures 10 and 11. } 
\label{fig11bis}
\end{figure}

\begin{figure*}
\psfig{figure=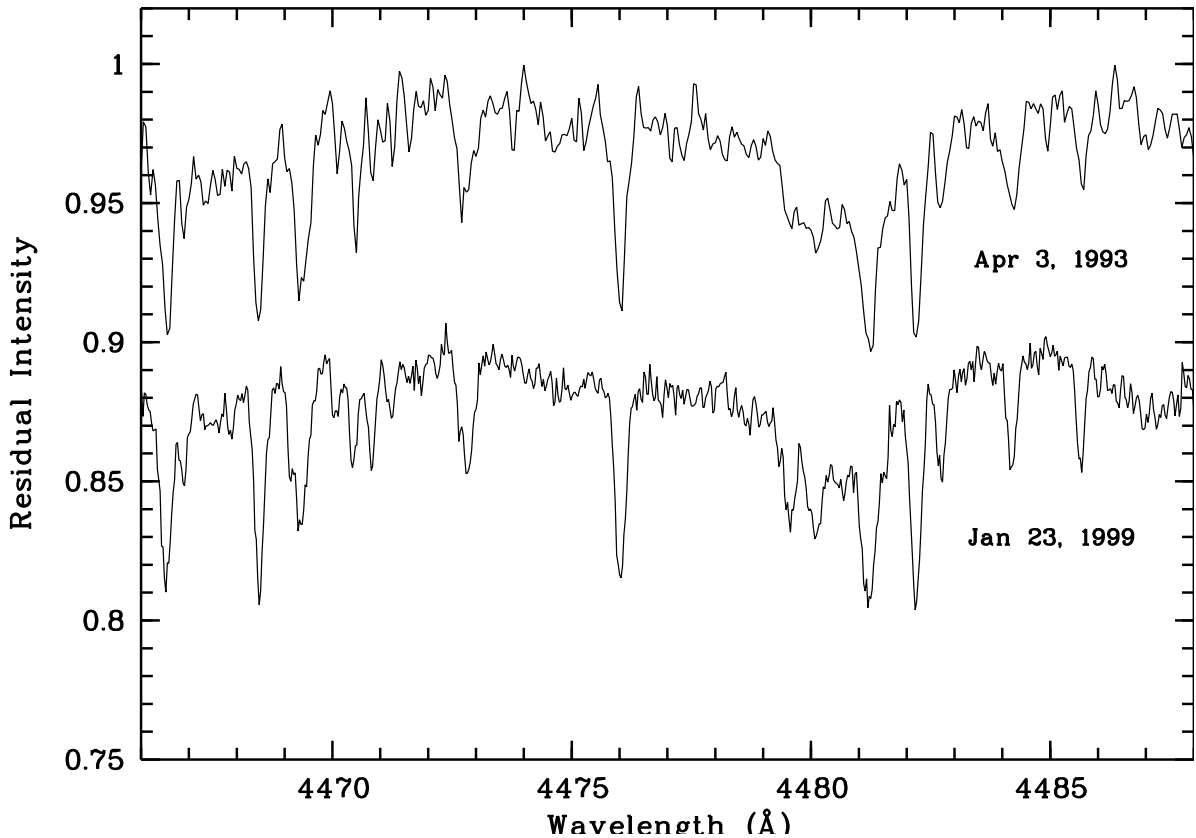,clip=t}
\caption{
Different profile shapes and asymmetries of narrow lines in the two 
spectra of HD 111786 observed at different epochs; the lower spectrum has 
been shifted by -0.1.
 }
\label{pl_fig11}
\end{figure*}

\begin{figure}
\psfig{figure=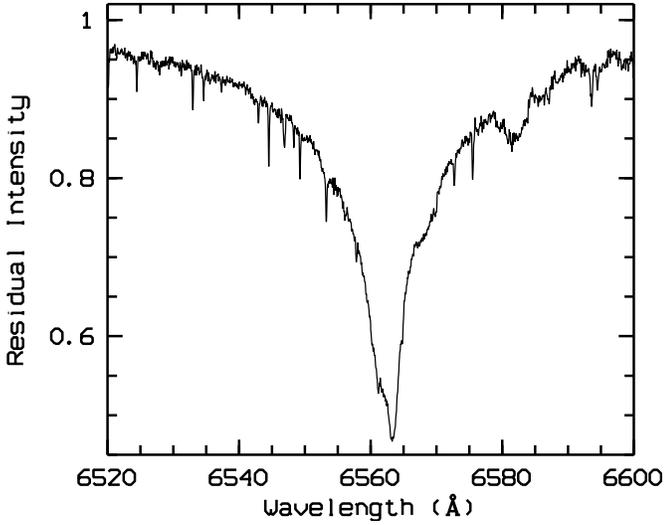,clip=t}
\caption{The distorted H$\alpha$ profile of HD 111786.} 
\label{fig12}
\end{figure}

\begin{figure}
\psfig{figure=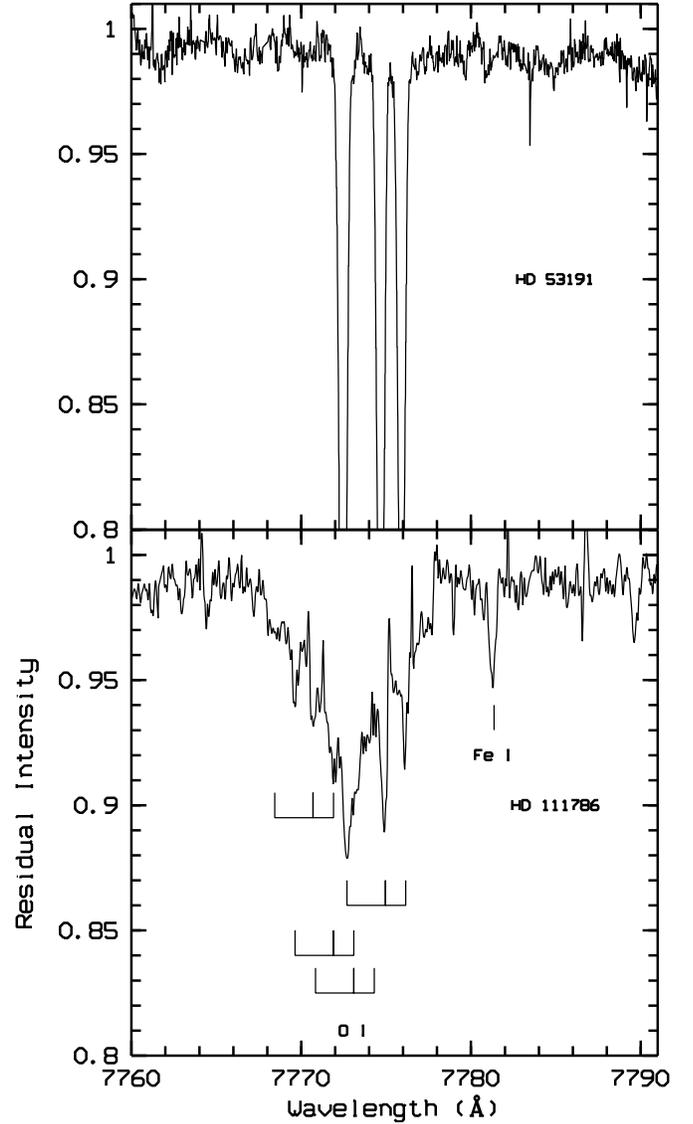,clip=t}
\caption{ The complex structure of the O I triplet 7772-7775 in the HD 111786
spectrum compared with the same feature in the low rotating A0 V star HD 53191.}
\label{fig13}
\end{figure}

\begin{figure}
\psfig{figure=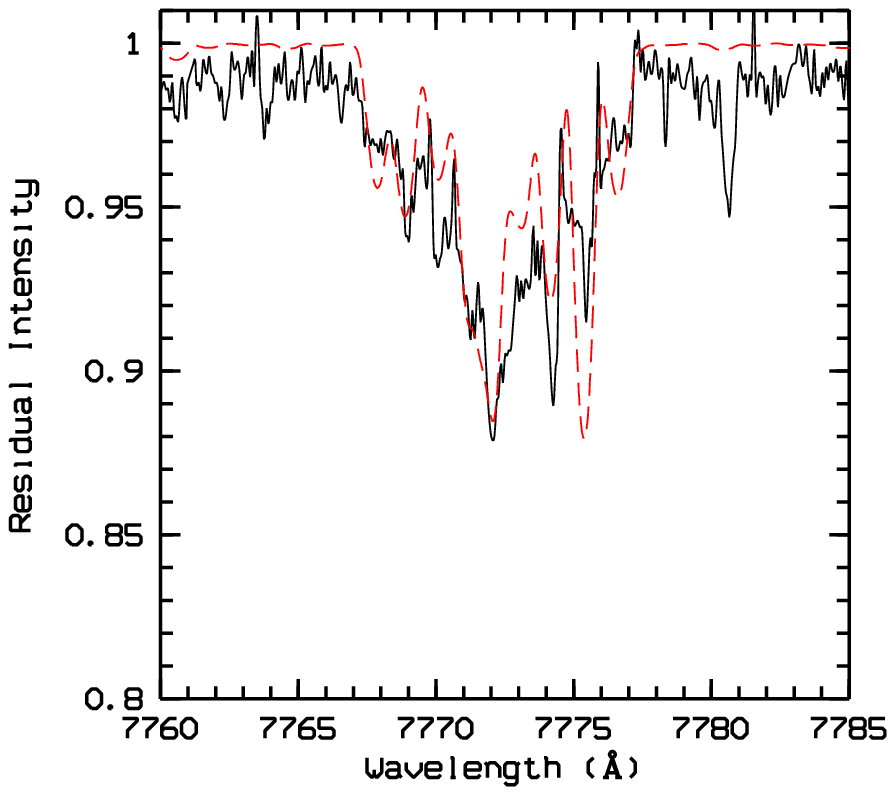,clip=t}
\caption{ The complex structure of the O I triplet 7772-7775 in the HD 111786
spectrum compared with that obtained by combining 4 synthetic spectra of slowly
rotating stars (dashed line); see text for details.}
\label{fig14}
\end{figure}

\subsection{Echelec data}
The  Echelec spectrograph was mounted at the 
Cassegrain focus of the ESO 1.52 m telescope at La Silla;
our spectra  cover the 4200-4500 \AA ~spectral range; 
the S/N ratio is not constant and highly 
variable between the centre and the edges of each echelle order.
The spectra  have been reduced with the package
written by Burnage \& Gerbaldi (1990 and 1992)  and running under
MIDAS.

\bigskip

\begin{table}
\caption{The observations}
\begin{center}
\begin{tabular}{ccrcc}
\hline
HD     &      J.D.    & S/N & Instrument & range (\AA)\\
\hline
 18622 & 2447789.8875 & 100 & Echelec & 4200-4500\\
 18622 & 2448524.8156 & 100 & Echelec & 4200-4500 \\
 18622 & 2448583.5424 &  60 & Echelec & 4200-4500 \\
 38545 & 2451621.3368 & 170 & Musicos & 5150-8800\\
 38545 & 2451622.3403 & 260 & Musicos & 5150-8800\\
 47152 & 2451622.2389 & 250 & Musicos & 5150-8800\\
 47152 & 2451623.3750 & 180 & Musicos & 5150-8800\\
 47152 & 2451817.6896 & 200 & Musicos & 5150-8800\\
 84948 & 2451622.4583 & 210 & Musicos & 5150-8800\\
111786 & 2449084.6875 & 150 & Echelec & 4200-4500\\
111786 & 2451201.8243 & 160 & FEROS   & 3800-8900\\
153808 & 2451622.6875 & 260 & Musicos & 5150-8800\\
153808 & 2451822.3333 & 210 & Musicos & 5150-8800\\
\hline
\end{tabular}
\end{center}
  \label{ Table 2
         }

\end{table}

\subsection{FEROS data}
 
The HD 111786 spectrum has been taken 
with the new high resolution spectrograph
FEROS (Kaufer et al. 1999) installed at the Coud\'e focus
of the ESO 1.5m telescope at La Silla.
This spectrograph is a bench-mounted fiber-fed echelle spectrograph. It is
installed in a
thermally stabilized, humidity controlled room.
The spectra are spread over 39 orders. The on-line
DRS package running under MIDAS performed the bias subtraction, the
flatfielding,
the wavelength calibration, the optimal extraction of the orders and
the merging of the orders.

\subsection{Musicos data}

The Musicos spectrograph is mounted at the 
2.2m Telescope Bernard Lyot of the
Observatoire du  Pic du Midi ; this echelle spectrograph  has a fiber entrance 
corresponding to 2.1 arcsec on the sky. Through two different prisms it 
may cover the full optical range in two exposures covering the 
blue and red spectral ranges respectively, each containing 46 orders.
We have only used the setting covering the 5150-8890 \AA~ region. 
The photometric and wavelength reduction have been done 
directly at the telescope site by using ESpRIT (Echelle Spectra 
Reduction: an Interactive Tool), a computer code for 
on-line processing developed by Donati et al (1997) and made available at the
telescope. 
 
\subsection{IUE data}

We inspected also 
IUE low and high resolution 
spectra extracted from the INES Archive.
We used the low-resolution and the high-resolution rebinned images for the
study of the flux distribution.
The high-resolution spectra have been examined only
for the echelle
orders where shell lines are expected.

\section{The spectra and their interpretation}

The aim of this work is not to make any abundance analysis of these binaries, 
but only to detect and interpret the differences between the observed 
spectra and those
computed with the hypothesis that these objects are single.
The features more sensitive to these differences will be selected to be used 
to search other binaries among the  \lambo candidates.

The whole available spectral range for each star has been examined since the
features most sensitive to duplicity are not necessarily the same 
for all objects; they depend on the atmospheric parameters and the 
magnitude difference of the binary system components. However, the wavelength
regions that revealed a peculiar behaviour in most stars are the Balmer line 
cores, the O I 7774 triplet and the Na I 5890-5896 doublet.

We shall describe in detail for each of the 5 chosen binaries the signs of
duplicity we have detected in the spectra at our disposal.

\subsection{Chosen atmospheric parameters for the computed spectra}

The determination of \teff and log~g
is obtained from the visual photometric colour indices by using the calibrations
of Moon and Dworetsky (1985 (MD)) and K\"unzli et al (1997) of the 
uvby${\beta}$ and Geneva photometric 
colour indices respectively and by assuming  the unrealistic hypothesis that 
the observed spectrum is produced by a single source.

The MD calibration is made for normal stars, but is valid for Ap stars too as
discussed by Hubrig et al (2000, Appendix A); 
however, the extension to the low-blanketed spectra of \lambo stars make 
sometimes doubtful the 
assignement to the groups defined in this calibration program. For
example, St\"urenburg (1993) [St93] chose the group 5 for 
all the 
stars of his sample, giving the priority to the spectral type A0-A3 of
the stars he analyzed. 
On the contrary, we give the highest 
priority to the ${\beta}$ index which is expected to be less 
sensitive to metallic lines;
in fact the spectral classification can easily assign
a too early spectral type to A stars with weak metal lines. 
Other criteria 
adopted by us are: the lowest resulting colour 
excess and the agreement with the parameters derived by other methods. 

The K\"unzli et al (1997) calibration is used in the hypothesis that 
the colour excess
is negligible as indicated by the Moon (1985) program results.

The parameters so derived are given in Table 3; the reddening is computed 
with the Moon (1985) program UVBYBETA and the $v \sin i$ values are taken from
the literature. 

The synthetic spectra are computed  with Kurucz program SYNTHE and 
by using his line-blanketed LTE 
models  Kurucz (1993).  The composite spectra have been 
computed either with the Kurucz
BINARY program or by combining the single synthetic spectra within MIDAS.
Broadening velocities are adopted in order
to fit the observed line widths.

\bigskip

\begin{table}
\caption{Photometrically derived atmospheric parameters; the $v \sin i $ values are 
from 
1) Slettebak et al (1975); 2) Abt \& Morrell (1995); 3) Paunzen et al (1998).
                              }
\begin{center}
\begin{tabular}{crccccc}
\hline

HD    &   E(b-y)& \teff &log g & \teff & log g & $v \sin i $  \\

      &    Moon &   MD  &  MD &   Gen &  Gen &  \kms \\   
\hline

18622 &  0.000  & 8070  & 3.60    & 8048  & 3.47      & 60(1)\\

38545 &  -0.030 & 8600  & 3.59    & 8495  & 3.58      & 175(2)\\

47152 &  0.010  & 10190 & 4.22    & 10021 & 4.23      & 25(2)\\

84948 &  -0.030 & 6780  & 3.39    & --    & --        & 45,55(3)\\

111786 &  0.000  & 7490  & 3.95    & 7397  & 4.29      & 135(2)\\

153808 &  0.000  & 10130 & 4.20    & 9987  & 4.20      & 50(2)\\

\hline
\end{tabular}
\end{center}
  \label{ Table 3
         }

\end{table}

\subsection{HD 18622} 

This SB2 binary is chosen as an example of a "normal" star; 
its spectral variations   are displayed
in Fig. \ref{fig1} and illustrate how 
a single spectrum  
may be not sufficient to exclude the duplicity of a star. 
For this star, the very similar parameters of the two components make
almost single the spectrum at the phase in which the lines
are not resolved 
(see Fig. \ref{fig2}). This Figure
shows the comparison between this observed spectrum and that computed with the
parameters given in Table 3, but with $v \sin i $=75 \kms , value 
slightly higher than
that given by Slettebak et al (1975). Even in this 
phase, where the star looks single, we note that several metal
lines are stronger in the synthetic 
spectrum, so
simulating globally a slightly weak-line  star; in particular it 
must be noted that in the
H$_{\gamma}$ profile,  the core is flatter than that computed,
contrarily to what is expected and observed in single stars. 
The core of the Balmer lines is very sensitive to NLTE effect.
When computed  in LTE, using the Kurucz code, it should be less deep than the
observed one, contrary to what is observed in Fig. \ref{fig2}.

The case of the normal HD~18622 
demonstrates that it may be impossible, at certain phases, to detect the
duplicity of an object without a careful comparison of observed with 
computed Balmer line profiles

\subsection{HD 38545}

The two spectra we examined are characterized by narrow components of several
strong lines, in particular of H$_{\alpha}$, of the Na I doublet and of
several lines of Fe II.

The absence of narrow components on Mg II 
4481 (see figure 11 in St93) and on the high excitation Fe II lines
indicates that the narrow lines are due to a shell, 
not to a companion star, which would have affected these lines too.

The presence of these shell components prevents us from performing a 
detailed comparison
between the observed spectrum and the two we have computed, one adopting 
the parameters chosen by us and listed in Table 3 and solar abundances, and 
the other with the St93 parameters and the metal abundances derived by this
author. 
The two H$_{\alpha}$ profiles computed as for a single star, 
either by using the 
parameters given in Table 3 or with those adopted by  St93 
(\teff=8970 K, log g=3.960, $v \sin i $=200 \kms ) do not reproduce 
well the 
observed one (Fig. \ref{fig3}), however it is impossible to decide to what 
extent this is due to the duplicity and to what extent this is due to the 
presence of the shell.

The triplet structure of the O I 7772-7775 is washed out by the high 
$v \sin i $ 
of this composite spectrum; it appears as a single absorption 
feature; a simple abundance change in each  of the two models would be
sufficient to reproduce observations;  
so this OI 7772-7775 triplet
is not a useful criterium to detect duplicity when its structure is masked by 
a high broadening.

This rapidly rotating star
represents the class of objects for which the spectroscopic approach 
alone is not sufficient to prove the  stellar duplicity.
The comparison with synthetic 
spectra can only provide more or less convincing indications on duplicity
from the inconsistencies between observations and computations.

\subsection{HD 47152}

HD 47152 (53 Aur) has been classified \lambo by Abt \& Morrell (1995), 
while it has been studied in the past as Ap of type  Hg-Eu-Cr.
The Catalano \& Renson (1998) catalogue of Ap and Bp stars 
reports a photometric period of 2.80 d and Adelman \& Pyper (1993) assign
this star to the group of those with moderate to no spectral variations.
The detection of the duplicity goes back to 1982 (McAlister \& Hendry, 1982)
and has been followed by 15 other papers on speckle measurements.
However, the effect of the companion on the observed spectrum has never 
been taken
into account in all the studies of this star, probably because these
speckle data do not give information on the magnitude difference between
the components of the binary system.
The Hipparcos Catalogue data, giving a magnitude difference of
${\Delta}$Hp=0.77, clearly indicate that  the observed spectrum must be 
affected by the presence of the companion star. 

The two spectra we have taken show that the star is a spectroscopic variable;
the Mg I triplet, plotted in Fig. \ref{fig5}, for example, is weaker on 
the second spectrum taken 1 day after the first one.

The spectral lines are narrow, so confirming the $v \sin i $ measured by
Abt \& Morrell (1995), however they are not fitted by the synthetic
spectrum computed with the parameters given in Table 3 (but with 
$v \sin i$=30 \kms )  and by assuming solar abundances. 
The wings of the H$_{\alpha}$ profile are reasonably 
well fitted by the computed spectrum ( Fig. \ref{fig6}),  but we note 
that the observed core is considerably less deep than that of the computed 
profile, contrary to what expected from NLTE
effects which are not accounted for in our computations.  The metal
lines appear not to be weaker than expected for a star with these values
of \teff and log g so the \lambo classification appears not to be
justified for this star, at least at the phases of our observations. 
In the contrary,  the high number of stronger than computed and of 
unidentified lines are in favour of the classification as Ap; however 
we remark some anomalies compared to classical Ap stars.
The OI triplet at 7774 A is a very broad feature (Fig. \ref{fig7}) which may 
be fitted by a 
spectrum computed with $v \sin i $=300 \kms ; we recall that the 
only previous 
mention of such a high $v \sin i $ is in Palmer et al (1968) where a value of 
325 \kms ~ is given for this star.
 Moreover this feature does not indicate the oxygen underabundance
characteristic of Ap stars (see also Gerbaldi et al (1989) systematic 
study of this feature in Ap stars).
We are not aware of any paper on Ap stars pointing out the discovery of
 non identified lines in the OI 7774 region.

The Na I doublet is very weak for the \teff of 10000 K,  the HeI 5876 
is absent and  the Fe I and 
Fe II lines would suggest a lower \teff, not coherent with the H$_{\alpha}$ 
profile.  In fact several Fe II lines appear very weak compared with
predictions (e.g. 5166.033, 5197.577, 5316.615 \AA), while the Fe I lines
are systematically stronger than the computed ones.
So several contradictory results obtained from the optical 
spectrum appearance suggest that HD 47152 is not a single object.
We can also add that in the UV the h and k lines of Mg II and the 
strong Fe II
lines in the range 2500-2600 \AA~ have a broad profile, not coherent with most
line profiles in the optical range. 

Considering the high inclination (i=119.5$^o$) and the relatively long
period (P=38.90 yrs) given in the IAU Inf Circ. Comm 26, No.142, 2000,
it is likely that radial velocity variations
may appear over a span of several years.

\subsection{HD 84948}

Moderate underabundances have been found by Paunzen et al (1998)
for both components of this SB2 system. We adopted the same atmospheric 
parameters used by these authors to reproduce the spectrum at our disposal,
i.e. \teff=6600 and 6800 K, log~g=3.3 and 3.7, $v \sin i $=45 and 55 
\kms ~ for each component respectively.
The corresponding models have been computed by 
adopting the  metal abundances derived by these authors and we computed the
synthetic spectra with the same quoted value of the microturbulence 
(v$_{turb}$=3.5 \kms ~ for both components) and adopting  
the hypothesis of these authors that the two stars have almost the same 
luminosity.
The observed  H$_{\alpha}$ line is used to derive the relative radial 
velocity of the components at the epoch of our observation: the values
we found are -77 and +70 \kms  respectively.
The resulting combined synthetic spectrum 
is displayed in Fig. \ref{fig8}  
superimposed on the observed one. 
We note that while the redshifted component is well
fitted, the reverse is true for the other component; any 
small change in the
luminosity ratio will worsen the fit of the redshifted component and does not
produce more convincing results for metal line profiles.
The fit of the NaI doublet (Fig. \ref{fig9}) {and of the OI 7774 triplet 
(Fig. 9)}  confirm that the parameters for the higher
rotating star do not reproduce the observations; they may even suggest that 
this component is indeed a double star.   
Also the other metal lines are not fitted by this computed spectrum, so that 
we cannot confirm the abundance values
reported in the above quoted paper.

This object shows that it is not easy to produce a combined spectrum of an SB2
star that matches the observed one because many free parameters have to be
taken into account and so it may happen that in a narrow wavelength range 
the observed spectrum 
is roughly fitted by a computed one, even if the adopted parameters are 
not the real ones. 

Moreover this kind of analysis does not have any sense if made on one or 
few spectra; in fact a careful check that the combined spectrum is not 
variable with the period is necessary to exclude the partial eclipse
possibility.

The bad fit with the combination proposed by Paunzen et al (1998) may be interpreted 
either as due to a choice of parameters that must be revised,  
as it can be expected from the inconsistency of the difference in log g of
the two components 
and the adopted almost equal luminosities, or as sign 
of variable spectrum.

\subsection{HD 153808}

The two spectra at our disposal, even if the  
one taken on October 4th has a lower S/N and
is affected by strong H$_{2}$O telluric bands,
clearly confirm that the star is an SB2; a more subtle check of the 
March 18th 
spectrum made through the comparison with a synthetic spectrum computed
from the model with \teff and log g given in Table 3, but with a lower 
value of the broadening (30 \kms ), reveals that the H$_{\alpha}$ profile 
(Fig. \ref{fig10}) of the  blue-shifted component is distorted, suggesting
that it is probably due to two sources.
This duplicity is supported by the complex structure of the O I triplet at
7774 A (Fig. 12) and more evident in Fig. \ref{fig11} where the region 
of the Na I doublet and He I 5876 is compared with the same computed spectrum. 

The cooler component is responsible for the strong and red-shifted 
component of the NaI doublet and for the FeI 5914 line, 
but does not contribute to the He I 5876 absorption. 
The hotter blue-shifted (on the plotted spectrum) component 
of the NaI doublet presents a double absorption which has the same
profile and separation as that of the HeI 5876 line. These profiles 
are interpreted as a signature (together with the H$_{\alpha}$ core 
profile) of duplicity of the hotter component of this system.

The considerable difference in the atmospheric parameters makes the star
brighter in the red that appears fainter in the photographic blue range 
examined by Petrie (1939) and
the high resolution of Musicos allowed us to detect the presence of a third 
component. 

Lambert et al (1986) by adopting \teff=9700 K, log g=4.00 and 
$v \sin i$ =38 \kms
~ derived log${\epsilon}(C)$=7.68 from lines at $\lambda$=9000 \AA; 
this extreme C deficiency is discussed in detail by these authors and the 
possibility of a companion examined. 

The present study clarifies that this object is a triple
spectroscopic system and not a \lambo star.

This star is an example of spectroscopic triple system with a too small angular 
separation of the components to be detected by Hipparcos (ESA, 1997) or by
speckle  interferometry (Marchetti et al, 2001).  
This example demonstrates that a single most efficient approach to detect 
all the \lambo binaries does not exist, since we cannot make  {\it a priori}
any guess on the angular separation  of the possible components.

\section{Binary detection criteria from spectrum inspection}

The fact that newly detected binaries have been found among stars already well
studied indicates that the two components of these systems have similar 
parameters; this fact combined with the quite high value of $v \sin i $, 
characteristic of A-type stars (not belonging to the Ap-Am classes) stars, makes 
it very delicate to detect the 
duplicity and an abundance analysis based on the average atmospheric 
parameters may be done by lowering metal abundances in order to compensate
the veiling effect. If the two components of a binary system are 
similar, the average spectrum is expected to be similar to that of each
component; this is the case of HD 38545 for which only mild underabundances
have been derived (St93). A larger difference of the
two sources is expected to produce larger apparent abundance peculiarities
as in the case of HD 47152, classified either Ap or \lambo.

We have shown that even a triple system can be confused with a single peculiar
star when the anomalous line intensities are attributed only to abundance
anomalies, as in the case of  HD 153808.

If one allows all  the abundances to be fitting parameters,
spectral synthesis will provide a reasonable fit to the data
in almost all cases, even though the fitting 
abundances are totally unphysical.
In the previous section we have shown that
a very careful inspection of 
any observed spectrum is mandatory before starting any abundance analysis
of an object that has a mean-high $v \sin i $ value.
We have shown that this is not an easy task and we have selected the
spectral features that are the more promising for A-type stars and that 
may be different for different binaries.

The example of HD 18622 has been given to underline the importance to have
more than a single spectrum; an SB2 may not be detectable at certain phases.

From the inspection of the other spectra we learned that for rapidly rotating 
stars, most of the absorption lines are in fact blends and the less 
contaminated features of  species present over a large interval of stellar
\teff are the best suited for such detection. We have selected in this way the 
O I 7774 triplet as the most helpful reference for stars with a moderate value
of $v \sin i $:   see figures  6, 9 and 12 .

The absorption core of the Balmer lines which is broadened mainly by 
Doppler effect is expected to be deeper than that computed in the LTE 
approach while it is flatter in composite spectra. But the intrinsic 
difficulty to trace a very accurate continuum 
in the region of Balmer lines for dwarf A-type stars spectra must not be
neglected, due to 
the fact that the maximum extension of the wings of these lines appear 
in the \teff range around  8500 K. 

The detection of unidentified 
lines must be carefully checked with some
reference star spectrum; in fact that may simply indicate a displaced
absorption by a companion star;
the presence of these lines 
is easier to detect at long wavelength where the line crowding is
less important.

The period of spectroscopic binaries is usually short; so more than 1 
spectrum taken  during an observing run should be sufficient to detect 
not very rapidly rotating SB2 stars.

It should be noted that we have not discussed in detail the comparison 
between the
observed and the computed flux distribution; in principle this can be 
a further criterion to recognize binaries if the two components have not equal
\teff. In practice, this comparison is rarely efficient because based, 
for most stars, on photometric magnitudes and the calibration of these values
are affected by uncertainties of the same order that the differences we
are looking for.

The discrepancies we have found in making these 
comparisons could be explained either by the duplicity or for the above
reasons and therefore we consider this approach a valid method only if 
relative comparisons are made between objects observed with the same
photometric filters for visual and IR.

When this comparison is restricted to the UV range covered by IUE spectra
the main sources of uncertainty are represented by the accuracy of the
computed line blanketing and by the r\^ole of the NLTE on the metal 
b-f discontinuities. 
The comparison in this range requires accurate computations of the
heavy line blanketing and of the b-f transitions. In fact these can be 
significantly
different in real metal deficient stars as the \lambo are supposed to be
and in stars with solar abundances.

So, in general, only complementary information can be obtained for suspected
complex objects when peculiarities are found in the UV range not coherent 
with what is expected from the visual range analysis.
This is the case with HD 38545 whose UV flux suggests a lower \teff or a 
higher blanketing than the optical spectral analysis and of HD 111786 (the 
star we shall discuss in the next Section) whose high UV flux suggests a  higher
\teff or a lower blanketing than that indicated by the optical spectrum.

\bigskip
\section{Application to the case of
HD 111786: a complex object, not a \lambo star}

HD 111786 (HR 4881) has been assigned to \lambo class by Andersen 
and Nordstr\"om (1977) 
and since then accepted by all the following authors.
The Hipparcos Catalogue (ESA, 1997) gives its parallax (16.62 $\pm$ 0.72 mas),
and no sign of duplicity is mentioned. 

Faraggiana et al (1997) discovered the duplicity of this "\lambo star" on the 
basis of the double core of the H$_{\gamma}$ profile and of the
set of narrow lines superimposed on the broad features in the examined 
optical spectrum (covering the limited wavelength range of 300 \AA~ from 4200 
to 4500 \AA). 
They showed also that a combination of two theoretical spectra both 
computed with 
{\it solar abundances}, \teff = 7500 K, log g = 4.0, but with different 
$v \sin i $ (10 and 150 \kms ) 
simulates qualitatively the observed spectrum. 
Moreover the high UV flux and
the absence of the narrow line components at wavelength shorter than 
2000 \AA~ suggested that the 
broad component was slightly hotter than the other.

On the basis of these indications, we made 
many attempts to reproduce the observed composite spectrum
by combining two theoretical spectra;  a rough, but not good fit was 
obtained
for the spectrum covering the 4200-4500 \AA~ range, by combining two spectra
computed from Kurucz models with solar abundances, log g=4 for both 
components, \teff=7750 and 7200 K, $v \sin i $=250 and 20 \kms ,
relative luminosities=0.7 and 0.3 and a radial velocity difference
of 50 \kms .  A real good fit has never been achieved by introducing  
slightly different parameters and various differences in the luminosity
ratio of the two components
and we were left with the doubt that the system is more complex than a simple
binary.

The experience gained by the analysis of the binaries discussed in previous 
sections suggested to examine a spectrum taken with higher resolution 
and covering a much broader wavelength 
domain than that previously used.

Contrary to HD 38545, in HD 111786 the narrow lines cannot be due to the 
presence of a shell; in fact all the 
spectral lines present the narrow components, not only those 
characteristic of stellar shells (i.e.  the resonance or low excitation
lines and those 
arising from a metastable 
level from which the atom cannot cascade to a lower level in a permitted
transition); even Mg II 4481 (lower E.P.=8.83 eV) presents the complex profile
corresponding to broad and narrow components.
The broad line component is the predominant one and that on which the
abundance analysis  by St\"urenburg (1993) is based; see also Fig 12 of his
paper. The complex and variable profiles of the narrow lines denote their
complex origin which is due to more than one low rotating source (Fig. 13). 

All the Balmer lines present a splitting of the core, 
but if a simple 
duplicity could have been derived from the H$_{\delta}$ double core,
the more complex multiplicity of this object appears from H$_{\alpha}$
whose profile shows the contribution of several components. The increasing
contribution of cooler components of this system appears also from the
increasing complexity of metal lines features toward long wavelength.
The profile of the O I 
triplet 7772-7775 which lies in a spectral region not contaminated by other
lines contribution is plotted in Fig. \ref{fig13} together with the same region 
observed in the low rotating A0~V HD 53191. 
The complex feature observed in HD 111786 is  
composed by a broad blend due to a high 
rotating star ($v \sin i $ at least 150 \kms ) (in the not proved 
hypothesis that 
this profile component is due to a single source)  and by 4 or more other 
narrow sources, one of which is slightly brighter than the others. 
We have produced the computed spectrum plotted in Fig. 16 by combining
a synthetic spectrum computed with the model \teff=7250 K with three others
with \teff=7750 K shifted by -157.0, -118.5, +46.0 \kms  with respect 
to the first one; all these spectra have been computed by assuming a broadening
of 20 \kms ~ and therefore the dominant broad-lined component is not included.
We do not claim that this combination is a faithful reproduction of the 
observed spectrum, but we only want to demonstrate how such a complex 
morphology may indeed arise due to stellar multiplicity.

The two examples in figures 15 and 16 show the complexity 
of this system and the existence of multiple distinct solutions.
So, the analysis of the profile of the OI 7774 triplet in particular,
clarified the problem
of the high multiplicity of this star.

HD 111786 appears to be formed by a clump of at least 5 stars and 
therefore we cannot hope 
to make any reliable abundance analysis of this object revealing possible 
abundance anomalies. If we accept the definition of \lambo
stars as single objects with peculiar atmospheric abundances then {\it the 
classification of this complex object as a \lambo star must be definitely
rejected}.

\section{Conclusion}

Faraggiana \& Bonifacio  (1999) suggested that binaries can be 
easily confused with \lambo stars 
when their classification is based only on classification dispersion spectra. 
The detection of binaries with the spectrum affected by a companion,
i.e.  of objects producing a composite spectrum, is the fundamental
first step before making any abundance analysis on which more or less
elaborated theories to explain these stars will be based.
However, the detection of binaries among these stars is subtle; the systems 
already detected show that the components of each system have a slight difference
in stellar parameters so that an abundance analysis based on the average 
value of the atmospheric parameters does not produce absurd results.

Most of the double systems that have been already recognized as such 
have an angular separation lower than 1 arcsec, i.e. below the 
limit of most spectrographs entrance aperture.
Most \lambo stars have mean-high $v \sin i $ and the detection of duplicity 
on the basis of spectra only remains often doubtful.

As a consequence several complementary approaches must be considered; 
the most promising techniques are speckle, 
adaptive optics and interferometry, but still a large gap in the 
angular separation is expected to remain between the limits of interferometric
and spectroscopic detection.
  
In the present paper we used the last approach and presented the results of 
search for criteria that can reveal the stellar duplicity from the inspection
of high quality spectra covering a large wavelength range.
Isolated  features, as the OI 7774 triplet, the NaI doublet, the core of 
the Balmer lines, 
and the general fit between observed and computed profiles appear to fulfill
our purpose. 

This study allowed us to recognize that HD 153808 is a triple
spectroscopic system.

The application of these selected criteria to a \lambo candidate, HD 111786,
revealed the complex nature of this object, which is not a single peculiar
star surrounded by a shell as presented in the literature, but a clump 
of at least 5 components.

\bigskip

\begin{acknowledgements}
Use was made of the SIMBAD data base, operated at the CDS,
Strasbourg, France.
RF acknowledges grants from MURST 40$\%$ and 60$\%$ and from the University
of Trieste.

\end{acknowledgements}

\end{document}